\documentclass[9pt,journal]{IEEEtran}
\usepackage{graphicx}
\usepackage{xspace}
\usepackage{xurl}
\usepackage{multirow}
\usepackage{multicol}
\usepackage{booktabs}
\usepackage[table,xcdraw]{xcolor}
\usepackage[hidelinks]{hyperref}

\newcommand{\mypara}[1]{\vspace{5pt}\noindent{\textbf{{#1}:}\xspace}}

\title{Bao-Enclave: Virtualization-based Enclaves for Arm}

\author{%
{\rm Samuel Pereira, João Sousa, Sandro Pinto, José Martins, David Cerdeira}\\ 
{\small Centro ALGORITMI, Universidade do Minho} \\ 
{\small	\{a81408, a82273\}@alunos.uminho.pt, \{sandro.pinto, jose.martins, david.cerdeira\}@dei.uminho.pt} \\
} % end author

\begin{document}
\maketitle    

\begin{abstract}
General-purpose operating systems (GPOS), such as Linux, encompass several million lines of code. Statistically, a larger code base inevitably leads to a higher number of potential vulnerabilities and inherently a more vulnerable system. To minimize the impact of vulnerabilities in GPOS, it has become common to implement security-sensitive programs outside the domain of the GPOS, i.e., in a Trusted Execution Environment (TEE). Arm TrustZone is the de-facto technology for implementing TEEs in Arm devices. However, over the last decade, TEEs have been successfully attacked hundreds of times. Unfortunately, these attacks have been possible due to the presence of several architectural and implementation flaws in TrustZone-based TEEs. In this paper, we propose Bao-Enclave, a virtualization-based solution that enables OEMs to remove security functionality from the TEE and move them into normal world isolated environments, protected from potentially malicious OSes, in the form of lightweight virtual machines (VMs). We evaluate Bao-Enclave on real hardware platforms and find out that Bao-Enclave may improve the performance of security-sensitive workloads by up to 4.8x, while significantly simplifying the TEE software TCB.

\end{abstract}

\section{Introduction}
General-purpose operating systems (GPOS), such as Linux, are nowadays significantly complex, encompassing several million lines of code. As all products of the human intellect, software is intrinsically subject to defects and statistically likely to present unexpected behaviors, generally referred to as "bugs". Thus, a larger code base inevitably leads to a higher number of potential vulnerabilities and inherently a more vulnerable system \cite{7163052}. These vulnerabilities can be leveraged to tamper with security-critical information and consequently subvert the Confidentiality, Integrity, and Availability (CIA) triad guarantees. To address this problem, it has become common to implement security sensitive programs outside the domain of the GPOS, running in Trusted Execution Environments (TEEs) \cite{Costan2016, Costan2016-2, Pinto2019}. TEEs provide an isolated execution environment, which the main OS cannot tamper with, used to protect the privacy and data integrity of applications (even under a compromised main system). Two of the most well-established TEE technologies are Intel Software Guard Extensions (SGX) \cite{Costan2016} and Arm TrustZone \cite{Pinto2019}, in the cloud/server and mobile/embedded domains, respectively. Both technologies aim at achieving similar high-level goals, but their architecture and implementation are significantly different.

Arm TrustZone \cite{Pinto2019, Pinto2019-virt} enforces the separation based on the concept of worlds, i.e., the normal world and the secure world. The secure world (a.k.a. TEE) is used for the security critical services, while the normal world for everything else, i.e., the GPOS and applications - the Rich Execution Environment (REE) \cite{Pinto2019, Cerdeira2020, Jang2018}. These worlds have separate dedicated memory regions and different privileges, and the secure monitor is responsible for switching between secure and non-secure execution. While the secure world software is able to access all normal world resources, the reverse is not possible. 
In the Intel SGX \cite{Costan2016, intel}, protected address areas (a.k.a. enclaves) are created for applications, which enforces protection at the hardware and software level. Enclaves aims at providing confidentiality and integrity even when the entire system software is compromised, as the enclave's memory region cannot be accessed by any software that does not reside in the same enclave \cite{8360317}.
While in the TrustZone architecture the trusted OS can access and tamper with trusted applications (TAs), in SGX, the main OS is not granted such privilege. In SGX, there is an extra (and intra-) level of isolation which, in theory, makes enclaves inherently more secure. Thus, the content of an enclave is protected and cannot be accessed by any process outside of the enclave perimeter, including processes executing at higher privilege levels. This fundamentally minimizes and contains the impact of potentially vulnerable enclave code.

Arm remains the main provider for computing platforms in the mobile world, with over 95\% \cite{arm95} of the world's smartphones market share. Although ARM platforms are prevalent, TrustZone-based systems have been show vulnerable hundreds of times \cite{Cerdeira2020}. The expressiveness of the problem encompasses the overall architecture, the software stack implementation, and the hardware. These flaws can lead to bypass the full TEE security and allow, for example, attackers to obtain secrets such as cryptographic keys and biometric authentication \cite{Cerdeira2020, Keegan2019}. Architecture deficiencies in the TEE system are due to fundamental design aspects of TrustZone such as the over-privileged rights of the secure world \cite{Cerdeira2020, Cerdeira2022}. Implementation bugs involves classic input validation errors, such as buffer overflows. At the hardware level, important hardware properties are overlooked at the architectural and microarchitectural level \cite{Cerdeira2020}.

To address mainly the aforementioned architectural TrustZone TEEs limitations, the academic community has proposed several solutions ~\cite{Cerdeira2022, vTZ, Brasser2019, tfence, Yun2019, Sun2015, havmsi}. Most of these solutions are built on hardware mechanisms, including virtualization techniques, which enforce access control based on the current executing context, while removing functionality from the TrustZone TEE. However, these solutions suffer from problems such as controlled channel attacks, require compiler changes, and suspension of the entire operating system while a trusted application (TA) executes, just to name a few. Additionally, other solutions have been proposed by academia~\cite{overshadow, inkTag, TrustVisor, usingHypervisor}  for commodity x86 platforms, which mainly use Intel virtualization technology. They feature a hypervisor, secure context switching, secure processor creation, and provide security mechanisms to protected applications.

%Virtualization stands as an appealing solution for implementing TEEs. It allows for strong isolation and introspection of common OSs, such as Virtual Machines (VMs)~\cite{5504800}. Virtualization can be used to restrict the memory accessible by the OS, which can be leveraged to create an environment to run applications that the OS cannot access. However, there are no solutions that leverage virtualization for TEE creating in ARM platforms.

In this paper, we present Bao-Enclave, a virtualization based solution to create enclaves on ARM(v8-A) platforms. The enclaves execute in lightweight virtual machines (VMs) in the normal world and allow developers to relocate complex functionality from the secure world, reducing the system TCB. Bao-Enclave is built atop the open-source static partitioning hypervisor Bao \cite{bao}, which we modify to support the dynamic creation of VMs. We evaluate Bao-Enclave on two well-established hardware platforms (NXP i.MX8MQ and Xilinx ZCU104) and compared with OP-TEE. The results are encouraging, achieving up to 4.8x better execution times.

\section{Background}

\subsection{TEEs on ARM platforms}
TrustZone is used for deploying TEEs \cite{Pinto2019, Oliveira2022}, which isolate the execution of security critical programs named Trusted Applications (TAs). Some TAs implement services for the operating system (OS), e.g., for user authentication or file disk encryption~\cite{androidwhitepaper}. Other TAs provide shared user-level functionality, e.g., DRM media decoders~\cite{TZVideoPath} or online banking services~\cite{trustonicbanking}. These TEEs comprise a stack that offers an API and runtime support for hosting TAs. TEEs such as Qualcomm's QSEE and OP-TEE protect the confidentiality and integrity of TAs' memory state, thereby ensuring it cannot be inspected or tampered with by a potentially compromised OS.

\begin{figure}
    \centering
    \includegraphics[width=0.7\linewidth]{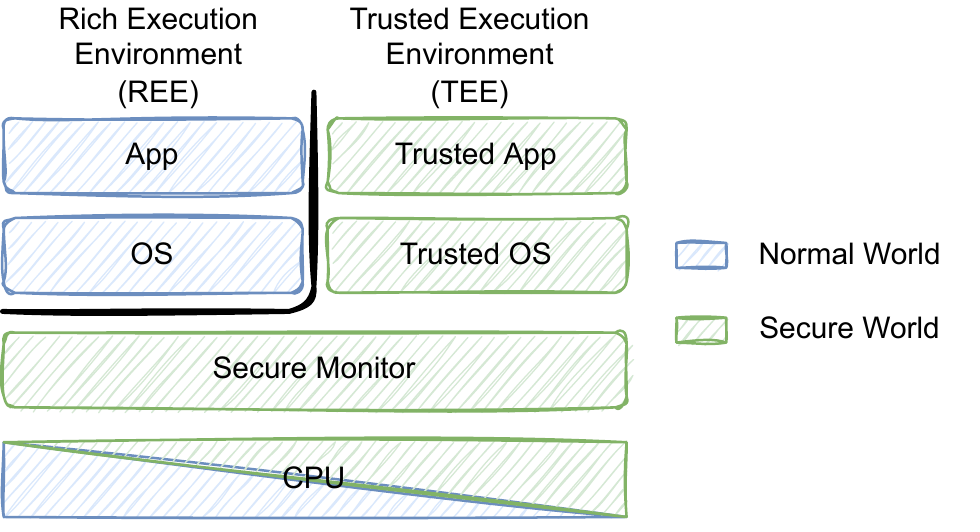}
    \caption{Arm TrustZone secure world executing the TEE, and normal world executing the REE.}
    \label{fig:armtz}
\end{figure}

TrustZone provides two execution environments: \textit{normal world} and \textit{secure world}. Figure~\ref{fig:armtz} depicts the normal world running a feature-rich software stack, named Rich Execution Environment (REE), which comprises a full-blown OS and applications, software considered untrusted. The secure world runs a simpler software stack comprised by a trusted OS and TAs. TrustZone enforces isolation between worlds and provides specific entry points for world switching, managed by the secure monitor, whenever the REE invokes TAs' services. A physical CPU can be in either of two security states, secure and non-secure, depending on which software it is executing. It is secure for monitor Trusted OS and TA, and non-secure for OS and applications. 

With ARM platforms being so prevalent, TrustZone-based systems have been show vulnerable multiple times.
The main security issues in TrustZone are derived from implementation issues that are amplified by the TrustZone's own architecture flaws~\cite{Cerdeira2020}.
Implementation bugs both in TAs and the Trusted OS, are typical implementation errors such as buffer overflows or dereferencing attacker controlled pointers. These flaws reach critical security impact due to the TrustZone architecture giving high privileges to TEE software. This results in TEE components that can be controlled by an attacker, and from which devastating attacks can be launched from affecting every part of the system~\cite{Cerdeira2020}. This is particularly worrisome as OEM strive for evermore complex functionality in the secure world leading to larger TCBs, which in its turn leads to less trustworthy systems.

\subsection{Virtualization}
% Virtualization is a technology capable of running multiple guests simultaneously, by partitioning just one physical machine (host) into multiple guests fully isolated temporally and logically from each other. Providing to the guests decoupling from the real hardware, setting up mutual isolation between guests.
% Virtualization only checks the execution of software that is authorized and operations of VMs are independent in the system except when VMs must synchronize their operation to cooperate on a task.
% Virtualization assumes the role of providing software with a virtual version of a given physical resource. Thus, this layer helps to solve a myriad of TEE problems.
% Trusted application code runs in a high-security. All code that does not belong to the same safe region is unable to perform any action on the safe zone memory. It is not possible to read the data efficiently because the addresses given at these privilege levels are only virtual addresses, a view of the physical addresses.

System virtualization~\cite{smithnair} is a technology that enables the consolidation of multiple, unrelated software stacks onto the same physical machine by partitioning and multiplexing hardware resources (e.g. CPUs, memory, etc) between multiple virtual machines (VMs). The software layer that implements virtualization is called virtual machine monitor (VMM) or hypervisor. The software executing in the VM is referred to as guest, typically an operating system, thus guest OS.

In their seminal paper, Popek and Goldberg~\cite{popekgoldberg} have identified the three main properties of virtualization as being (i) equivalence, meaning that the provided VM abstraction should be as compatible as possible to that of the physically machine so that guest software can run unmodified; (ii) efficiency, which implies there should be minimal performance deterioration of guest software as compared to when executing directly on the real machine; and, more importantly, (iii) resource control. In essence, resource control means that all VMs must be thoroughly isolated from one another. All system resources must be in full control of the hypervisor which delegates them to VMs, and it must be impossible for a VM to access resources affected to other VMs, unless explicitly configured so.
Furthermore, they have developed a model that lays out the basic requirements for a machine to be virtualizable. It essentially states that besides providing at least two privilege modes of execution and memory isolation mechanisms (e.g. page-based virtual memory), an instruction set must make sure that all sensitive instructions (i.e. instructions that either configure or depend on the configuration of the system) must also be privileged instructions which when executed in a lower privilege mode, trap to the higher privilege modes. The hypervisor will run in a high privilege mode, while guest software executes in a lower privilege. The hypervisor can than react to the traps originated due to the a guest executing sensitive instructions (a.k.a. VM-exits) by emulating the behavior of the physical machine while guaranteeing the resource control property, e.g. by implementing shadow page-tables~\cite{adams} to guarantee spatial isolation among all VMs.

Due to its current widespread applicability, architectures have introduced virtualization extensions~\cite{uhlig, varanasi, Sa2022}, which simplifies the implementation of hypervisors and reduce overheads by minimizing the frequency of VM-exists. They typically provide an extra privilege mode for the hypervisor itself, support for two-stage virtual memory translation, and fine-grained controls for trapping guest execution.

Starting from the 1960s, virtualization technology has been used to time-share a single physical machine among multiple users~\cite{smithnair, garfinkle2005}. In the beginning of the century, due to the World Wide Web explosion, this technology has been extensively adopted in server environments to consolidate the many workloads, achieving higher utilization and lessening power consumption~\cite{garfinkle2005, stoess2007} thereby decreasing total data center maintenance costs. More recently, the technology has found its way into many other domains of application such as mobile and embedded (e.g. automotive)~\cite{virt_embedded, kaiser2009}, not only due to its cost saving benefits, but also by enabling significantly smaller form factors, the decoupling of the software stack from the real hardware easing sharing, maintenance, upgradability and portability, and, more importantly, due to its security and isolation benefits. As virtualization guarantees a high degree of isolation between VMs, it can be used to decompose the system, following the principle of the least privilege~\cite{saltzer75} which can greatly improve fault containment. This allows system designers to segregate security-sensitive functionality in dedicated VMs, essentially implementing software-defined TEEs. Furthermore, the hypervisor is a suitable layer to implement security functionality such as monitoring mechanisms~\cite{garfinkel2003}.

\subsection{Bao Hypervisor}\label{sec:bao}

Bao~\cite{bao} is a multi-core embedded hypervisor targeting mixed-criticality use cases where typically small safety- or mission-critical RTOSs or baremetal applications run in VMs alongside VMs hosting feature-rich guest OSs. In this domain, the main goal of the hypervisor is to help consolidate these multiple workloads while guaranteeing thorough isolation and strong real-time guarantees. Also, the hypervisor must minimize attack surface and vectors, but also be suitable for certification. To accomplish these goals, Bao implements a static partitioning architecture~\cite{li2011quest,ramsauer2017look}, meaning all system resources are allocated and assigned to VMs at initialization time and never reconfigured during execution. Virtual CPUs (vCPUs) are mapped to physical CPUs (pCPUs) in a 1:1 manner and IO is purely passthrough, therefore without the need for a scheduler. In this way, Bao is able to achieve a TCB of about 8.4 KSLoC for the Armv8 architecture while being a completely standalone implementation not depending on any external libraries. Bao implements inter-VM communication based on statically defined shared memory and asynchronous events mapped as virtual interrupts in the VMs communicating through a given channel.
Bao also enables cache partitioning via page coloring~\cite{liedtke} reducing the contention in shared caches and improving predictability and determinism for critical workloads. This partitioning can also mitigate cache-based timing side-channels used in a myriad of modern attacks~\cite{lyu2018survey}.

\mypara{VM Stacking} Despite simplicity and minimality being pillars in Bao's design, they are also one of its main drawbacks as the static and exclusive assignment of CPUs sacrifices utilization for determinism even when it is not necessary. To circumvent this, an experimental Bao branch implements a policy-free mechanism that relaxes the static 1:1 mapping of vCPUs to pCPUs, allowing multiple vCPUs belonging to different VMs to execute on the same pCPU. However, the hypervisor only implements a simple context-switch mechanism and remains without a scheduler. Instead, at configuration time, it is possible define a tree of vCPUs for each pCPU. At runtime, the executing vCPU can issue a hypercall to schedule any of its child vCPUs. A vCPU can also issue a hypercall to yield execution to its parent vCPU. If at any time of a child vCPU execution, an interrupt arrives targeting a parent vCPU, the hypervisor will immediately schedule it. Furthermore, other exceptions unhandled by the hypervisor are also forwarded the parent vCPU. This mechanism is dubbed VM stacking as the scheduling and yielding of child vCPU can be seen as pushing and popping it of the stack, respectively. Also, the scheduling of a parent vCPU upon interrupt arrival can be seen as stack unwinding since this can result in popping multiple vCPUs from the stack until the target vCPU is reached.

Essentially, VM stacking allows the implementation of arbitrarily complex functionality without increasing the TCB of critical VMs by moving it to a a high privilege VM, i.e., a VM in one of the root nodes of the configuration tree. 
The most straightforward example would be to implement scheduling itself in one of the CPUs: the root node vCPU would decide which of its child VMs according to some policy at each timer tick. Note this does not have any effect on a critical VM executing alone in another CPU.

\section{Motivation, Design Goals and Threat Model}

Bao-Enclave focuses on the secure processing of critical data on Armv8 platforms. The security critical applications must not trust the OS, as the OS can be compromised by attackers. The typical solution would be to run this security critical code in the TEE. However, if the application is flawed, it can lead to a full system compromise~\cite{Cerdeira2020}. Thus, this work's main goal is to provide a solution that enables the creation of safe execution environments for security critical applications in the normal world, while preventing these security critical applications from compromising the system. To this end, Bao-Enclave will use one primary VM to host the main OS. This VM will be able to request the creation of additional VM destined to host TAs inside an enclave, enclave VMs. Additionally, Bao-Enclave will provide an API and development models similar to SGX's to allow for flexible enclave management and deployment. 

The TCB of a running Bao-Enclave includes all secure world components, as they retain the highest privilege, and the Bao hypervisor. A TA hosted in an enclave does not directly need to trust the OS or any other normal world component, except for Bao, as they do not have privileges to access the enclave's code and data.

\section{Design and Implementation}

\begin{figure}
    \centering
    \includegraphics[width=0.55\linewidth]{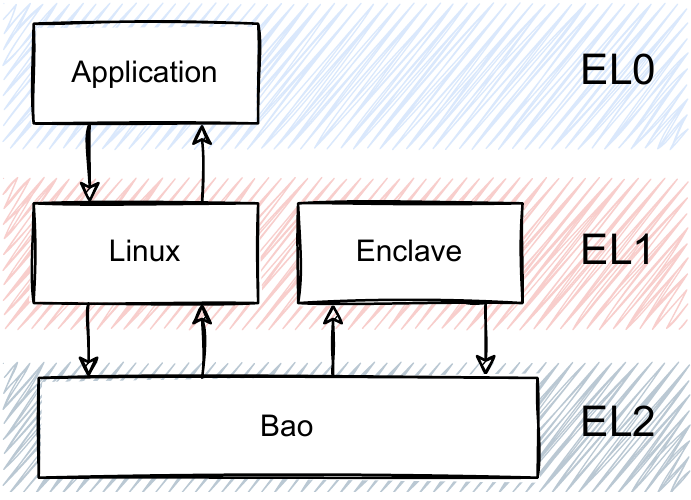}
    \caption{Bao-Enclave architecture overview.}
    \label{fig:Bao-Enclave_arch}
\end{figure}

Bao-Enclave creates isolated memory regions for security applications, hereinafter referred to as TAs. The memory region is provided as part of a VM which includes both EL1 and EL0 privilege levels. A TA developer can opt to build its application as a baremetal app and deploy it in EL1, build it's application with a library OS, or deploy a more typical software stack including an OS and run a TA, or multiple TAs, on top of it. For simplicity, we depict in Figure~\ref{fig:Bao-Enclave_arch} the scenario in which a TA runs as a baremetal application in EL1.
To create an isolated environment in the normal world, we use virtualization techniques. Specifically, we dynamically create enclave VMs, while having a primary VM hosting the main OS. We use stage-2 page tables to control physical memory access. This is needed to prevent the primary VM (i.e., the OS) from accessing, or otherwise compromise, the enclave VMs.
Bao-Enclave leverages the Bao hypervisor due to its small size, and because it provides strong isolation between VMs. As Bao is a static partitioning hypervisor, it does not feature the ability to dynamically create guest VMs. Therefore, a small number of modifications were made to Bao to implement this functionality, including the implementation of the hypercall interface necessary to allow the primary VM to request the creation of enclaves. In its original form, Bao allows a single VM to run in one physical CPU. We take advantage of a work-in-progress feature in Bao, called VM Stack, see section~\ref{sec:bao}.

Bao-Enclave follows the same general model as SGX, applying it to ARM processors, but with some differences.
%memory encryption
First, Bao-Enclave does not protect against hardware attacks such as memory bus snooping~\cite{bussnooping}, or cold boot~\cite{coldboot}. SGX achieves this by having a memory encryption engine in the SoC that encrypts and decrypts data on the fly when data is being sent to, or fetched from, main memory, respectively.
However, Bao-Enclave could be extended to provide similar features by implementing a paging mechanism and using on-chip memory~\cite{softme, sectee}.
% enclave creation
Enclave creation in SGX is done by the OS by using initialization instructions specific for enclave creation. These instructions inform the hardware which memory regions belong to the enclave, and mark the enclave as initialized.
In Bao-Enclave, the OS issues calls to the hypervisor to create and initialize an enclave. The OS must donate memory from it's own address space, and pass additional information to Bao in order to create the enclave.
% Context switching
One last significant difference between SGX and Bao-Enclave, is that SGX applications have a CPU instruction available to them, to invoke a TA. In Bao-Enclave an application must issue a call to the Bao-Enclave driver to perform a call to the corresponding TA.

% -------------------------------------------------------------------------- %
\subsection{Physical Address Space control}
\begin{figure}
    \centering
    \includegraphics[width=0.7\linewidth]{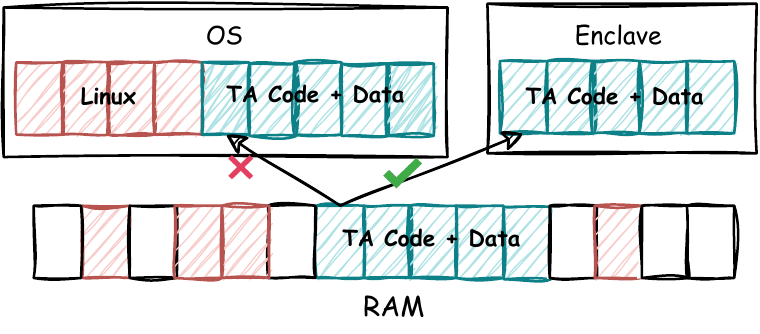}
    \caption{Physical memory access control by controlling stage-2 translation tables.}
    \label{fig:remocao}
\end{figure}

The main objective of Bao-Enclave is to create an isolated environment that protects security sensitive applications from being compromised by a malicious OS. The main mechanism Bao-Enclave uses to achieve this is stage-2 translation tables (i.e., virtualization). During the enclave VM creation process, Bao removes access to the physical memory now belonging the enclave VM from the primary VM running the main OS. Figure~\ref{fig:remocao} depicts the result of applying this mechanism. Although Linux has mapped the TA code and data unto it's own address space, the stage-2 translation table blocks access to this memory, allowing only the enclave VM to access it.

% -------------------------------------------------------------------------- %

\subsection{Bao-Enclave Execution Flow}
Bao-Enclave requires interactions between a user-space application and a Bao-Enclave driver on the host OS, and between the driver and the Bao hypervisor. These interactions have three main objectives: Enclave Creation, Enclave Destruction, and Enclave Invocation. A communication protocol is implemented at the application and TA level, to establish a connection between them.

\begin{figure}
    \centering
    \includegraphics[width=0.7\linewidth]{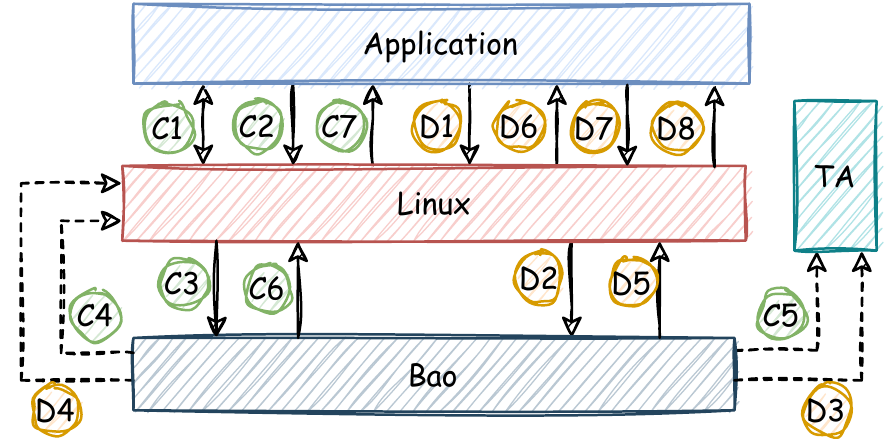}
    \caption{Enclave creation (C1-C7) and destruction (D1-D8) execution flow.}
    \label{fig:enclave_c/d}
\end{figure}

\mypara{Enclave Creation}
The creation process is depicted in Figure~\ref{fig:enclave_c/d} in steps C1 to C7.
When an application requires the execution of security critical code, it will issue a request to the OS through the Bao-Enclave driver (C1) to allocate memory for the enclave. The OS will then take some of its own memory and donate it to an enclave VM. This memory will hold the TA image (e.g., code and data) and be large enough for the TA to execute correctly. Both the TA image and the required memory space are previously established and stored in a file. The size of the communication channel (i.e., shared memory region) is also information stored in the file.
After the OS allocates the necessary resources, The application will copy the TA information to the newly allocated memory, and issue a request to create the enclave VM (C2). This request is first received by the Bao-Enclave driver, and then a similar request is sent to Bao (C3). Herein lies the most significant modification to Bao. The modifications transform Bao from being able to only create VMs during startup to being able to create them dynamically, specifically for the enclave use case. In this step Bao will take the memory region that the OS allocated to the enclave and remove it from the primary VM physical address space (C4), while mapping that same physical memory to the enclave VM (C5). After the enclave is fully created, Bao will give back execution control to the OS (C6), which will execute the application (C7), and the TA can then be invoked by the application.

% O hypervisor é o responsável pela criação da enclave. Por isso, o endereço onde a enclave se encontra no espaço de endereçamento do OS é passado para o hypervisor. No entanto, o código da própria enclave continua a pertencer ao OS. Isto é algo indesejável porque permite a exploração de ataques, tal como o acesso a dados privados da enclave. Afim de solucionar este problema, o hypervisor traduz os endereços GPA para endereços HPA, como representado na figura \ref{fig:traducao_os_hypervisor}, de forma a conseguir interpretar os endereços do OS.

% -------------------------------------------------------------------------- %

\mypara{Enclave Destruction} 
Destroying an enclave VM requires the execution of similar steps to its creation, but in reverse. These steps are represented in Figure~\ref{fig:enclave_c/d}, in steps D1 to D8. When an application no longer requires the TA services, it issues an enclave destruction call to the Bao-Enclave driver (D1). The OS will then issue a call to Bao to destroy the enclave (D2), to regain access to the memory it donated.  Bao will then destroy the VM (D3), another modification we introduce in Bao. In the destruction process Bao will write the enclave VM's memory region to zero, thus preventing the OS from learning secrets when it regains access to the memory region. After this, Bao will remap the memory region unto the primary VM's address space (D4), and control is given back to the OS (D5), and eventually the application (D6). The application will then issue a call to the OS to free the memory allocated for the enclave (D7). Finally the OS will give back control to the application (D8).

\begin{figure}
    \centering
    \includegraphics[width=0.82\linewidth]{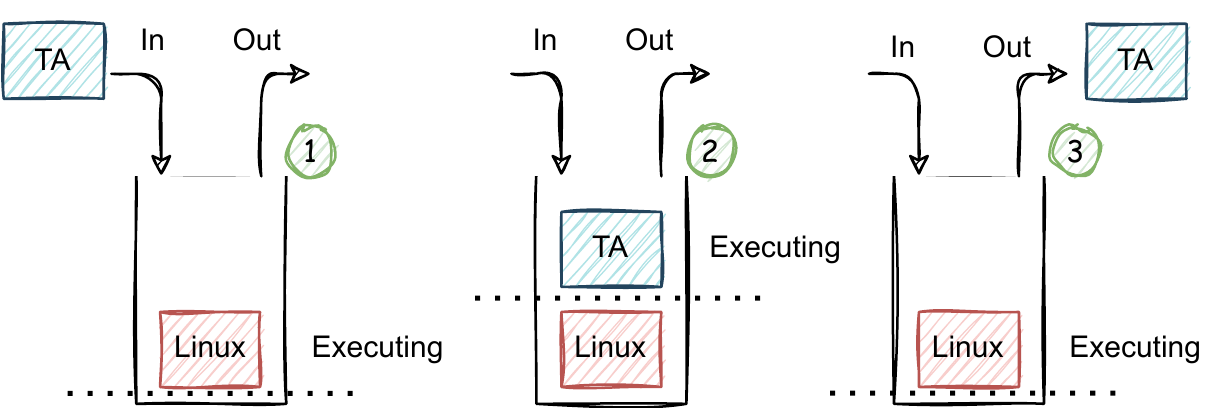}
    \caption{How a LIFO data structure is used to control the execution flow of multiple VMs on the same pCPU in Bao-Enclave.}
    \label{fig:lifo}
\end{figure}

% -------------------------------------------------------------------------- %
\newcommand{\Head}{\emph{HEAD}\xspace}
\newcommand{\Tail}{\emph{TAIL}\xspace}
\mypara{Invoking an Enclave}
To invoke the execution of an existing enclave, an application must issue a call to the Bao-Enclave driver. The driver will then issue a hypercall to Bao, which will perform the context switch. 
We leverage the two additional fields of the VM data structure, \Head and \Tail , introduced by Bao's VM Stacking feature. \Tail holds which VM executed pior and \Head stores the next VM to be invoked. 
%Upon handling a request the VM will return to Bao, and Bao will look at the \Tail field to determine the VM to execute, and update the VM data structure accordingly. 
This creates a LIFO data structure, used to keep track of the execution flow between VMs.
On receiving the invoke call, Bao will update the \Head field of the permanent VM data structure and set it to point to the enclave VM. After this, Bao will perform the context switch, and give execution control to the enclave VM hosting the TA. The TA will execute, and once it has processed the request it will issue a call to Bao to give control back to the OS. Because the TA is giving back control of the execution, Bao will look at the \Tail data field to retrieve the VM that invoked the TA. After this, control is given back to the OS. Figure~\ref{fig:lifo} demonstrates how the LIFO data structure is used to control the execution flow between VMs.

% -------------------------------------------------------------------------- %

\mypara{Communication}
Bao-Enclave inherits the shared memory based communication mechanism, present in the original Bao. However, contrary to the original Bao implementation, Bao-Enclave foregoes interrupt based communication, with the goal of simplifying the implementation. Thus, application and TA implement a shared communication protocol without relying on interrupts.
The TA must implement a loop, where execution resumes on an invoke request. In this loop the TA decodes which request it must serve. When the request is fulfilled, it gives back control, and the loop restarts.

\section{Performance Evaluation}
We evaluate Bao-Enclave using an i.MX8MQ from NXP and an ZCU104 from Xilinx, both platforms feature Cortex-A53 cores. While i.MX8MQ features a core frequency of up to 1.5GHz, a 32KiB L1 caches for data and instruction, and a 2MiB L2 cache, the ZCU104 features a core frequency of 1.2GHz, 32KiB L1 data and instructions caches, and 1MiB L2 cache. The toolchain we use is gcc-arm-10.2-2020.11. Our performance analysis covers three vectors, micro-benchmarks, real world use case, and TCB size.

\mypara{Micro-Benchmarks}
We first measure the elapsed time for the execution of each Bao-Enclave API: create enclave, invoke enclave, destroy enclave. We execute each test 30 times in the ZCU104 platform.
Table~\ref{tab:execucao_zcu} shows our results.
The operation of entering and exiting the enclave has a much smaller cost, 7µs, compared to the operations of creating and destroying the enclave. The added cost of the create and destroy operations is due to the required stage-2 page table maintenance for both the enclave VM and the primary VM.
The creation operation requires 138.58ms on average.
As for the enclave destruction, the cost is 1491.64ms, on average, which in addition to the page table maintenance similar to the creation operation, also includes the step of writing the enclave VM memory to zero.
\begin{table}[!t]
\caption{Bao-Enclave Micro-Benchmarks in ZCU104 platform.}
\label{tab:execucao_zcu}
\centering
\begin{tabular}{lccc}
                                        & Create  & Invoke & Destroy \\ \hline
{Average (ms)}         & 138.58 & 70.38{\footnotesize E}-4 & 1491.64  \\
{Std. Deviation (µs)} & 763.56 & 0.58   & 25.17   \\ \hline

\end{tabular}
\end{table}

\mypara{Real World Use case}
We evaluate the real world performance of a bitcoin TA running in OP-TEE, and a port of this TA running in a Bao-Enclave enclave. We measure 30 execution times for each of the commands provided by the TA for both the OP-TEE and the Bao-Enclave versions, and compare them to assess the incurred overheads.
The results are shown in Table~\ref{tab:comp_bao_tee}. The comparison is performed between an existing application for OP-TEE and a baremetal version implemented for this work. The TA is not same for OP-TEE and Bao-Enclave, as any function that uses libraries or calls to OP-TEE is replaced by a version implemented in C. As can be seen, Bao-Enclave presents better results than OP-TEE, in the comparison between all commands, with the exception of command~1 where Bao-Enclave is slower, this is caused by differences between applications. Apart from creating the master key where Bao-Enclave is about 8.2x slower. The main reason for the increased performance with Bao-Enclave is because several methods in the API used by the Bitcoin Wallet require system calls to OP-TEE, whereas all operations in the Bao-Enclave version of the TA are done in the same privilege level. We observe that the TA in Bao-Enclave can be up to 4.8x faster in the execution of the services. Finally, we observe that Bao-Enclave has a significantly lower standard deviation compared to the OP-TEE.

\begin{table}[!t]
\caption{Bitcoin Wallet execution times for Bao-Enclave and OPTEE in the i.Mx8MQ platform.}
\label{tab:comp_bao_tee}

\setlength{\tabcolsep}{3pt}
\centering
\scalebox{0.9}{
\begin{tabular}{llccc}

                       &            & Average (ms) & Relative Average & Std. Deviation (µs) \\ \hline
\multirow{2}{*}{Cmd 1} & Bao-Enclave & 20,9         &  8.4x            & 134,6               \\ 
                       & OP-TEE     & 2,5          &  ---             & 325,0               \\  \hline
\multirow{2}{*}{Cmd 2} & Bao-Enclave & 82,3         &  0.6x            & 64,2                \\ 
                       & OP-TEE     & 131,7        &  ---             & 5849,5              \\  \hline
\multirow{2}{*}{Cmd 3} & Bao-Enclave & 74,1         &  0.6x            & 21,9                \\ 
                       & OP-TEE     & 131,6        &  ---             & 4643,1              \\\hline
\multirow{2}{*}{Cmd 4} & Bao-Enclave & 17,3         &  0.9x            & 58,7                \\ 
                       & OP-TEE     & 19,7         &  ---             & 3554,5              \\ \hline
\multirow{2}{*}{Cmd 5} & Bao-Enclave & 59,3         &  0.2x            & 27,0                \\ 
                       & OP-TEE     & 285,8        &  ---             & 2786,1              \\  \hline
\multirow{2}{*}{Cmd 6} & Bao-Enclave & 51,8         &  0.3x            & 13,3                \\ 
                       & OP-TEE     & 194,7        &  ---             & 2020,9              \\ \hline
\end{tabular}
}
\end{table}

\mypara{TCB Size}
We measure the size of the TCB by evaluating how much lines of code are added, and how much the program segments are increased by the modifications necessary to implement Bao-Enclave. To measure the lines of code we use Tokei\footnote{https://github.com/XAMPPRocky/tokei}. To measure the application segment sizes we use the \emph{size} tool provided by the toolchain. We compare these values to typical Bao and to the Bao implementation including the VM Stacking mechanism. The results are presented in Table~\ref{tab:impacto_tcb}. Bao-Enclave performs all its functions with an addition of only 164 lines of code in the hypervisor, which represents a 1.79\% increase. These 164 lines represent changes to Bao's core C module, keeping the other sectors practically unchanged. Regarding memory usage, we analyze the size of the program segments in the binary. An increase in allocated memory is detectable, as expected. This increase corresponds to 1.27 KiB or 0.67\%.

\begin{table}[!t]
\caption{Bao-Enclave impact on system TCB.}
\label{tab:impacto_tcb}
\centering
\scalebox{0.8}{
\setlength\tabcolsep{2pt}
\begin{tabular}{c|ccccccccc|cllccll|}
\cline{2-17}
\multicolumn{1}{l|}{}                               & \multicolumn{9}{c|}{Lines of Code}                                                                                                                                                                                                                                                                                                                                                         & \multicolumn{7}{c|}{Size (bytes)}                                                                                                                                                                                                                              \\ \cline{2-17} 
                                                    & \multicolumn{3}{c|}{Bao}                                                                                                             & \multicolumn{3}{c|}{\begin{tabular}[c]{@{}c@{}}Bao \\ vmstack\end{tabular}}                                                          & \multicolumn{3}{c|}{\begin{tabular}[c]{@{}c@{}}Bao-Enclave\end{tabular}}                       & \multicolumn{3}{c|}{}                               & \multicolumn{1}{c|}{}                                                                         & \multicolumn{3}{c|}{}                                                                                       \\ \cline{2-10}
                                                    & C                            & \multicolumn{1}{c|}{asm}                         & \multicolumn{1}{c|}{\cellcolor[HTML]{D3D3D3}total} & C                            & \multicolumn{1}{c|}{asm}                         & \multicolumn{1}{c|}{\cellcolor[HTML]{D3D3D3}total} & C                            & \multicolumn{1}{c|}{asm}                         & \cellcolor[HTML]{D3D3D3}total & \multicolumn{3}{c|}{\multirow{-3}{*}{Bao}}          & \multicolumn{1}{c|}{\multirow{-3}{*}{\begin{tabular}[c]{@{}c@{}}Bao \\ vmstack\end{tabular}}} & \multicolumn{3}{c|}{\multirow{-3}{*}{\begin{tabular}[c]{@{}c@{}}Bao-Enclave\end{tabular}}} \\ \hline
\multicolumn{1}{|c|}{armv8}                         & 4864                         & \multicolumn{1}{c|}{514}                         & \multicolumn{1}{c|}{\cellcolor[HTML]{E6E6E6}5378}  & 5225                         & \multicolumn{1}{c|}{508}                         & \multicolumn{1}{c|}{\cellcolor[HTML]{E6E6E6}5733}  & 5209                         & \multicolumn{1}{c|}{508}                         & \cellcolor[HTML]{E6E6E6}5717  & \multicolumn{3}{c|}{125159}                         & \multicolumn{1}{c|}{128068}                                                                   & \multicolumn{3}{c|}{127628}                                                                                 \\
\multicolumn{1}{|c|}{imx8mq}                        & 45                           & \multicolumn{1}{c|}{0}                           & \multicolumn{1}{c|}{\cellcolor[HTML]{E6E6E6}45}    & 45                           & \multicolumn{1}{c|}{0}                           & \multicolumn{1}{c|}{\cellcolor[HTML]{E6E6E6}45}    & 45                           & \multicolumn{1}{c|}{0}                           & \cellcolor[HTML]{E6E6E6}45    & \multicolumn{3}{c|}{760}                            & \multicolumn{1}{c|}{776}                                                                      & \multicolumn{3}{c|}{776}                                                                                    \\
\multicolumn{1}{|c|}{core}                          & 2577                         & \multicolumn{1}{c|}{3}                           & \multicolumn{1}{c|}{\cellcolor[HTML]{E6E6E6}2580}  & 2849                         & \multicolumn{1}{c|}{3}                           & \multicolumn{1}{c|}{\cellcolor[HTML]{E6E6E6}2852}  & 3029                         & \multicolumn{1}{c|}{3}                           & \cellcolor[HTML]{E6E6E6}3032  & \multicolumn{3}{c|}{50841}                          & \multicolumn{1}{c|}{62347}                                                                    & \multicolumn{3}{c|}{64090}                                                                                  \\
\multicolumn{1}{|c|}{lib}                           & 495                          & \multicolumn{1}{c|}{0}                           & \multicolumn{1}{c|}{\cellcolor[HTML]{E6E6E6}495}   & 511                          & \multicolumn{1}{c|}{0}                           & \multicolumn{1}{c|}{\cellcolor[HTML]{E6E6E6}511}   & 511                          & \multicolumn{1}{c|}{0}                           & \cellcolor[HTML]{E6E6E6}511   & \multicolumn{3}{c|}{2828}                           & \multicolumn{1}{c|}{2812}                                                                     & \multicolumn{3}{c|}{2812}                                                                                   \\ \hline
\rowcolor[HTML]{D3D3D3} 
\multicolumn{1}{|c|}{\cellcolor[HTML]{D3D3D3}total} & \cellcolor[HTML]{E6E6E6}7981 & \multicolumn{1}{c|}{\cellcolor[HTML]{E6E6E6}517} & \multicolumn{1}{c|}{\cellcolor[HTML]{D3D3D3}8498}  & \cellcolor[HTML]{E6E6E6}8630 & \multicolumn{1}{c|}{\cellcolor[HTML]{E6E6E6}511} & \multicolumn{1}{c|}{\cellcolor[HTML]{D3D3D3}9141}  & \cellcolor[HTML]{E6E6E6}8794 & \multicolumn{1}{c|}{\cellcolor[HTML]{E6E6E6}511} & 9305                          & \multicolumn{3}{c|}{\cellcolor[HTML]{D3D3D3}179588} & \multicolumn{1}{c|}{\cellcolor[HTML]{D3D3D3}194003}                                           & \multicolumn{3}{c|}{\cellcolor[HTML]{D3D3D3}195306}                                                         \\ \hline
\end{tabular}
}
\end{table}

%\section{Security Analysis}

%\section{Discussion}
%\david{Future work}
%\david{Communication}

\section{Related Work}
The two main motivations for this work have already been addressed separately in two major lines of work, creating an environment isolated from the main OS without relying on TEE technology, and reducing the TCB in the secure world of TrustZone. 
Works addressing the first line of work include the use of virtualization techniques. Overshadow~\cite{overshadow} leverages shadow page tables to create different views of physical memory. Inktag~\cite{inkTag} introduces the concept of paraverification to improve isolation. TrustVisor~\cite{TrustVisor} implements TPM functionality in software that is leveraged as a root of trust for deploying secure VMs. SP3~\cite{sp3} encrypts VM memory by storing a per domain (i.e., collection of VMs) secret key in the page tables. TFence~\cite{tfence} leverages Armv7A partially privilege mode to securely instantiate and execute portions of applications, with the goal of guaranteeing a secure communication channel with the secure world. Bao-Enclave stands as the first solution aiming to reduce secure world TCB in Armv8 platforms while relying on a minimal, and fit for purpose, hypervisor.

% -------------------------------------------------------------------------- %

In the second line of work, mechanisms are implemented by the TEE to increase the security guarantees in the normal world, lessening the need to execute complex applications in the secure world. These works leverage the TZASC, a TrustZone address space controller that controls what memory is normal and which memory is secure. Ginseng~\cite{Ginseng} leverages TrustZone's secure monitor to implement a shadow stack that holds sensitive information available only while sensitive portions of the application execute. Sanctuary~\cite{Brasser2019} leverages non-standard features, including the TZASC-400 per master filtering abilities to create enclaves in the normal world. TrustICE~\cite{TrustICE} protects sensitive applications in the secure world while they are not executing, transitioning them to the normal world when they are needed. HA-VMSI~\cite{havmsi} and vTZ~\cite{vTZ} both leverage TrustZone to improve the security guarantees of a hypervisor, taking from the hypervisor its direct control over the address space of VMs. ReZone~\cite{Cerdeira2022} partitions the TEE in multiple domains by using COTS hardware features to create sandboxes in the secure world, effectively deprivileging the Trusted OSes. There are also works that use software techniques, such as same privilege isolation, to manage the TCB in the secure world~\cite{pros, teev}.
In Bao-Enclave we securely instantiate workloads previously hosted by the TEE to reduce the secure world TCB, without relying on TrustZone.

\section{Conclusion}
In this paper, we discussed Bao-Enclave, a virtualization based solution to create enclaves on ARM(v8-A) platforms. The enclaves execute in the normal world and allow developers to relocate complex functionality from the secure world, reducing the system TCB. Bao-Enclave is built atop Bao, which we modify to support the dynamic creation of VMs. We evaluate our system and compared it with OP-TEE. The results are encouraging, achieving up to 4.8x better execution times. In the future, we plan to add other features provided by SGX such as remote attestation and re-design our system to generalize to other computer architectures (e.g., RISC-V \cite{Sa2022}).

\section*{Acknowledgments}
% %-------------------------------------------------------------------------------
% 
We thank the reviewers for their comments and suggestions. This work was supported by national funds through Centro ALGORITMI / Universidade do Minho, under project UIDB/00319/2020. David Cerdeira and José Martins were supported by FCT grants SFRH/BD/146231/2019 and SFRH/BD/138660/2018 respectively.

%
%
% ---- Bibliography ----
%
% BibTeX users should specify bibliography style 'splncs04'.
% References will then be sorted and formatted in the correct style.
%
\bibliographystyle{unsrt}
\bibliography{references.bib}

\begin{thebibliography}{10}

\bibitem{7163052}
Y.~{Xu}, W.~{Cui}, and M.~{Peinado}.
\newblock {Controlled-Channel Attacks: Deterministic Side Channels for
  Untrusted Operating Systems}.
\newblock In {\em Proc. of S\&P}, 2015.

\bibitem{Costan2016}
Victor Costan and Srinivas Devadas.
\newblock {Intel SGX explained}.
\newblock In {\em Cryptology ePrint Archive}, 2016.

\bibitem{Costan2016-2}
Victor Costan, Ilia Lebedev, and Srinivas Devadas.
\newblock Sanctum: Minimal hardware extensions for strong software isolation.
\newblock In {\em Proc. of USENIX Security}, 2016.

\bibitem{Pinto2019}
Sandro Pinto and Nuno Santos.
\newblock {Demystifying Arm TrustZone: A Comprehensive Survey}.
\newblock {\em ACM Comput. Surv.}, 2019.

\bibitem{Pinto2019-virt}
Sandro Pinto, Hugo Araujo, Daniel Oliveira, Jose Martins, and Adriano Tavares.
\newblock {Virtualization on Trustzone-Enabled Microcontrollers? voil{\`a}!}
\newblock In {\em Real-Time and Embedded Technology and Applications Symposium
  (RTAS)}. IEEE, 2019.

\bibitem{Cerdeira2020}
D.~{Cerdeira}, N.~{Santos}, P.~{Fonseca}, and S.~{Pinto}.
\newblock {SoK: Understanding the Prevailing Security Vulnerabilities in
  TrustZone-assisted TEE Systems}.
\newblock In {\em Proc. of S\&P}, 2020.

\bibitem{Jang2018}
J.~{Jang}, C.~{Choi}, J.~{Lee}, N.~{Kwak}, S.~{Lee}, Y.~{Choi}, and B.~{Kang}.
\newblock {PrivateZone: Providing a Private Execution Environment Using ARM
  TrustZone}.
\newblock {\em IEEE Transactions on Dependable and Secure Computing}, 2018.

\bibitem{intel}
{Intel Software Guard Extensions}.
\newblock
  \url{https://software.intel.com/content/www/us/en/develop/topics/software-guard-extensions.html}.

\bibitem{8360317}
I.~{Sfyrakis} and T.~{Gross}.
\newblock {UniGuard: Protecting Unikernels Using Intel SGX}.
\newblock In {\em Proc. of Conference on Cloud Engineering (IC2E)}, 2018.

\bibitem{arm95}
{Kristin Bent}.
\newblock {ARM Snags 95 Percent Of Smartphone Market, Eyes New Areas For
  Growth}.
\newblock
  \url{https://www.crn.com/news/components-peripherals/240003811/arm-snags-95-percent-of-smartphone-market-eyes-new-areas-for-growth.htm},
  2012.

\bibitem{Keegan2019}
Keegan Ryan.
\newblock {Hardware-Backed Heist: Extracting ECDSA Keys from Qualcomm's
  TrustZone}.
\newblock In {\em ACM CCS}, 2019.

\bibitem{Cerdeira2022}
David Cerdeira, Jos{\'e} Martins, Nuno Santos, and Sandro Pinto.
\newblock {REZONE: Disarming TrustZone with TEE Privilege Reduction}.
\newblock In {\em Proc. of USENIX Security}, 2022.

\bibitem{vTZ}
Z.~Hua, J.~Gu, Y.~Xia, H.~Chen, B.~Zang, and H.~Guan.
\newblock {VTZ: Virtualizing ARM Trustzone}.
\newblock In {\em Proc. of USENIX Security}, 2017.

\bibitem{Brasser2019}
Ferdinand Brasser, David Gens, Patrick Jauernig, Ahmad-Reza Sadeghi, and
  Emmanuel Stapf.
\newblock {SANCTUARY: ARMing TrustZone with User-space Enclaves}.
\newblock 2019.

\bibitem{tfence}
J.~{Jang} and B.~B. {Kang}.
\newblock {Retrofitting the Partially Privileged Mode for TEE Communication
  Channel Protection}.
\newblock {\em IEEE Transactions on Dependable and Secure Computing}, 2020.

\bibitem{Yun2019}
Min~Hong Yun and Lin Zhong.
\newblock {Ginseng: Keeping Secrets in Registers When You Distrust the
  Operating System}.
\newblock 2019.

\bibitem{Sun2015}
{TrustICE: Hardware-Assisted Isolated Computing Environments on Mobile
  Devices}.
\newblock volume 2015-September, 2015.

\bibitem{havmsi}
Min Zhu, Bibo Tu, Wei Wei, and Dan Meng.
\newblock {HA-VMSI: A Lightweight Virtual Machine Isolation Approach with
  Commodity Hardware for ARM}.
\newblock 2017.

\bibitem{overshadow}
Xiaoxin Chen, Tal Garfinkel, E.~Christopher Lewis, Pratap Subrahmanyam, Carl~A.
  Waldspurger, Dan Boneh, Jeffrey Dwoskin, and Dan~R.K. Ports.
\newblock {Overshadow: A Virtualization-Based Approach to Retrofitting
  Protection in Commodity Operating Systems}.
\newblock ACM, 2008.

\bibitem{inkTag}
Owen Hofmann, Sangman Kim, Alan Dunn, Michael Lee, and Emmett Witchel.
\newblock {InkTag: Secure Applications on an Untrusted Operating System}.
\newblock {\em ASPLOS}, 2013.

\bibitem{TrustVisor}
Jonathan~M McCune, Yanlin Li, Ning Qu, Zongwei Zhou, Anupam Datta, Virgil
  Gligor, and Adrian Perrig.
\newblock {TrustVisor: Efficient TCB reduction and attestation}.
\newblock In {\em Proc. of S\&P}, 2010.

\bibitem{usingHypervisor}
Jisoo Yang and Kang~G. Shin.
\newblock {Using Hypervisor to Provide Data Secrecy for User Applications on a
  Per-Page Basis}.
\newblock In {\em Proc. of ACM SIGPLAN/SIGOPS International Conference on
  Virtual Execution Environments}. ACM, 2008.

\bibitem{bao}
José Martins, Adriano Tavares, Marco Solieri, Marko Bertogna, and Sandro
  Pinto.
\newblock {Bao: A Lightweight Static Partitioning Hypervisor for Modern
  Multi-Core Embedded Systems}.
\newblock In {\em {Workshop on Next Generation Real-Time Embedded Systems}},
  2020.

\bibitem{Oliveira2022}
Daniel Oliveira, Tiago Gomes, and Sandro Pinto.
\newblock {uTango: An Open-Source TEE for IoT Devices}.
\newblock {\em IEEE Access}, 2022.

\bibitem{androidwhitepaper}
{Google}.
\newblock {Android security white paper}.
\newblock
  \url{https://static.googleusercontent.com/media/1.9.24.55/en/US/work/android/files/android-for-work-security-white-paper.pdf},
  2015.

\bibitem{TZVideoPath}
{Michael Lu}.
\newblock {TrustZone, TEE and Trusted Video Path Implementation
  Considerations}.
\newblock
  \url{https://www.arm.com/files/event/Developer_Track_6_TrustZone_TEEs_and_Trusted_Video_Path_implementation_considerations.pdf},
  2018.

\bibitem{trustonicbanking}
{Trustonic}.
\newblock {Trustonic application protection delivers comprehensive security for
  mobile financial services}.
\newblock \url{https://www.trustonic.com/markets/financial-services/}, 2018.

\bibitem{smithnair}
J.E. Smith and Ravi Nair.
\newblock {The architecture of virtual machines}.
\newblock {\em Computer}, 2005.

\bibitem{popekgoldberg}
Gerald~J. Popek and Robert~P. Goldberg.
\newblock {Formal Requirements for Virtualizable Third Generation
  Architectures}.
\newblock {\em Commun. ACM}, 1974.

\bibitem{adams}
Keith Adams and Ole Agesen.
\newblock {A Comparison of Software and Hardware Techniques for X86
  Virtualization}.
\newblock ACM, 2006.

\bibitem{uhlig}
R.~Uhlig, G.~Neiger, D.~Rodgers, A.L. Santoni, F.C.M. Martins, A.V. Anderson,
  S.M. Bennett, A.~Kagi, F.H. Leung, and L.~Smith.
\newblock {Intel virtualization technology}.
\newblock {\em Computer}, 2005.

\bibitem{varanasi}
Prashant Varanasi and Gernot Heiser.
\newblock {Hardware-Supported Virtualization on ARM}.
\newblock ACM, 2011.

\bibitem{Sa2022}
Bruno Sá, José Martins, and Sandro Pinto.
\newblock {A First Look at RISC-V Virtualization From an Embedded Systems
  Perspective}.
\newblock {\em IEEE Transactions on Computers}, 2022.

\bibitem{garfinkle2005}
M.~Rosenblum and T.~Garfinkel.
\newblock {Virtual Machine Monitors: Current Technology and Future Trends}.
\newblock {\em Computer}, 2005.

\bibitem{stoess2007}
Jan Stoess, Christian Lang, and Frank Bellosa.
\newblock {Energy Management for Hypervisor-Based Virtual Machines}.
\newblock In {\em Proc. of USENIX ATC}, 2007.

\bibitem{virt_embedded}
Gernot Heiser.
\newblock {Virtualizing embedded systems-why bother?}
\newblock In {\em Design Automation Conference (DAC)}, 2011.

\bibitem{kaiser2009}
Robert Kaiser.
\newblock {Complex Embedded Systems - A Case for Virtualization}.
\newblock In {\em Workshop on Intelligent solutions in Embedded Systems}, 2009.

\bibitem{saltzer75}
J.H. Saltzer and M.D. Schroeder.
\newblock The protection of information in computer systems.
\newblock {\em Proceedings of the IEEE}, 1975.

\bibitem{garfinkel2003}
Tal Garfinkel, Mendel Rosenblum, et~al.
\newblock {A Virtual Machine Introspection Based Architecture for Intrusion
  Detection}.
\newblock In {\em Proc. of NDSS}, 2003.

\bibitem{li2011quest}
Ye~Li, Matthew Danish, and Richard West.
\newblock {Quest-V: A Virtualized Multikernel For High-Confidence Systems}.
\newblock {\em arXiv preprint}, 2011.

\bibitem{ramsauer2017look}
Ralf Ramsauer, Jan Kiszka, Daniel Lohmann, and Wolfgang Mauerer.
\newblock {Look mum, no VM exits!(almost)}.
\newblock {\em arXiv preprint}, 2017.

\bibitem{liedtke}
J.~Liedtke, H.~Hartig, and M.~Hohmuth.
\newblock {OS-controlled cache predictability for real-time systems}.
\newblock In {\em Proc. of IEEE Real-Time Technology and Applications}, 1997.

\bibitem{lyu2018survey}
Yangdi Lyu and Prabhat Mishra.
\newblock {A Survey of Side-Channel Attacks on Caches and Countermeasures}.
\newblock {\em Journal of Hardware and Systems Security}, 2018.

\bibitem{bussnooping}
Dayeol Lee, Dongha Jung, Ian~T Fang, Chia-Che Tsai, and Raluca~Ada Popa.
\newblock {An Off-Chip Attack on Hardware Enclaves via the Memory Bus}.
\newblock In {\em Proc. of USENIX Security}, 2020.

\bibitem{coldboot}
Sergei Skorobogatov.
\newblock {Low temperature data remanence in static RAM}.
\newblock Technical report, University of Cambridge, Computer Laboratory, 2002.

\bibitem{softme}
Meiyu Zhang, Qianying Zhang, Shijun Zhao, Zhiping Shi, and Yong Guan.
\newblock {Softme: A Software-Based Memory Protection Approach For TEE System
  to Resist Physical Attacks}.
\newblock {\em Security and Communication Networks}, 2019.

\bibitem{sectee}
Shijun Zhao, Qianying Zhang, Yu~Qin, Wei Feng, and Dengguo Feng.
\newblock {Sectee: A software-based approach to secure enclave architecture
  using tee}.
\newblock In {\em Proc. of ACM SIGSAC Conference on Computer and Communications
  Security}, 2019.

\bibitem{sp3}
J~Yang.
\newblock {Using Hypervisor to Provide Application Data Secrecy on a Per-Page
  Basis}.
\newblock In {\em In Proc. of Virtual Execution Environments}, 2008.

\bibitem{Ginseng}
Min~Hong Yun and Lin Zhong.
\newblock {Ginseng: Keeping Secrets in Registers When You Distrust the
  Operating System}.
\newblock In {\em Proc. of NDSS}, 2019.

\bibitem{TrustICE}
He~Sun, Kun Sun, Yuewu Wang, Jiwu Jing, and Haining Wang.
\newblock {Trustice: Hardware-Assisted Isolated Computing Environments on
  Mobile Devices}.
\newblock In {\em IEEE/IFIP International Conference on Dependable Systems and
  Networks}, 2015.

\bibitem{pros}
D.~{Kwon}, J.~{Seo}, Y.~{Cho}, B.~{Lee}, and Y.~{Paek}.
\newblock {PrOS: Light-weight Privatized Secure OSes in ARM TrustZone}.
\newblock {\em IEEE Transactions on Mobile Computing}, 2019.

\bibitem{teev}
Wenhao Li, Yubin Xia, Long Lu, Haibo Chen, and Binyu Zang.
\newblock {TEEv: Virtualizing Trusted Execution Environments on Mobile
  Platforms}.
\newblock In {\em ACM SIGPLAN/SIGOPS International Conference on Virtual
  Execution Environments}, 2019.

\end{thebibliography}

% \section{Meeting Notes}
% \begin{verbatim}
% - Estudo da shared memory
% - Estudo do VM stacking
% - Estudo da criação dinâmica de VMs
% - Estudo da criação dinâmica de VMs com VM stacking
    % * P: enclaves preparadas para correr em qualquer cpu? Para já fica tudo no mesmo
% \end{verbatim}
% 
%
% \begin{thebibliography}{8}
% \bibitem{ref_article1}
% Author, F.: Article title. Journal \textbf{2}(5), 99--110 (2016)
% 
% \bibitem{ref_lncs1}
% Author, F., Author, S.: Title of a proceedings paper. In: Editor,
% F., Editor, S. (eds.) CONFERENCE 2016, LNCS, vol. 9999, pp. 1--13.
% Springer, Heidelberg (2016). \doi{10.10007/1234567890}
% 
% \bibitem{ref_book1}
% Author, F., Author, S., Author, T.: Book title. 2nd edn. Publisher,
% Location (1999)
% 
% \bibitem{ref_proc1}
% Author, A.-B.: Contribution title. In: 9th International Proceedings
% on Proceedings, pp. 1--2. Publisher, Location (2010)
% 
% \bibitem{ref_url1}
% LNCS Homepage, \url{http://www.springer.com/lncs}. Last accessed 4
% Oct 2017
% \end{thebibliography}
\end{document}